\documentclass[sigconf,nonacm]{acmart}
\usepackage{algorithm}
\usepackage{algpseudocode}

\acmDOI{}
\acmISBN{}
\acmYear{2026}
\copyrightyear{2026}
\acmConference[FRecSys '26]
  {The 1st Workshop on Frontier Recommender Systems}
  {November 15, 2026}{Hong Kong, China}
\newcommand{\step}[1]{\ding{\numexpr171+#1\relax}}

\AtBeginDocument{%
  }

\begin{document}

\title{RecIS: Unifying Sparse and Dense Computation for Efficient Recommendation Training}
\subtitle{Author version. Accepted to FRecSys 2026
  (ACM SIGOPS ATC 2026).}

\makeatletter
\@ifundefined{correspondingauthor}{%
  \newcommand{\correspondingauthor}{}%
}{}
\makeatother

\author{Hua Zong}

\correspondingauthor

\authornote{
Both authors contributed equally to this work.
}

\affiliation{%
  \institution{Alibaba Group}
  \country{China}
}
\email{yuhuan.zh@alibaba-inc.com}

\author{Shuai Yang}

\authornotemark[2]

\affiliation{%
  \institution{University of Chinese Academy of Sciences}
  \country{China}
}
\email{yangshuai2022@iscas.ac.cn}

\author{Zhengxiong Zhou}
\affiliation{%
  \institution{Alibaba Group}
  \country{China}
}
\email{zhouzhengxiong.zzx@alibaba-inc.com}

\author{Qingtao Zeng}
\affiliation{%
  \institution{Alibaba Group}
  \country{China}
}
\email{zengqingtao.zqt@alibaba-inc.com}

\author{Jiawei Liu}
\affiliation{%
  \institution{Alibaba Group}
  \country{China}
}
\email{lanling.ljw@alibaba-inc.com}

\author{Hao Wan}
\affiliation{%
  \institution{Alibaba Group}
  \country{China}
}
\email{wanhao.wh@alibaba-inc.com}

\author{Hongbin Zhou}
\affiliation{%
  \institution{Alibaba Group}
  \country{China}
}
\email{hongbin.zhouhb@alibaba-inc.com}

\author{Jiahao Wang}
\affiliation{%
  \institution{Alibaba Group}
  \country{China}
}
\email{wjh177423@alibaba-inc.com}

\author{Hao Zhang}
\affiliation{%
  \institution{Alibaba Group}
  \country{China}
}
\email{zh138764@alibaba-inc.com}

\author{Bencheng Yan}
\affiliation{%
  \institution{Alibaba Group}
  \country{China}
}
\email{bencheng.ybc@taobao.com}

\author{Yuheng Huang}
\affiliation{%
  \institution{Alibaba Group}
  \country{China}
}
\email{hexing.hyh@taobao.com}

\author{Huimin Yi}
\affiliation{%
  \institution{Alibaba Group}
  \country{China}
}
\email{huimin.yhm@alibaba-inc.com}

\author{Jiamang Wang}
\affiliation{%
  \institution{Alibaba Group}
  \country{China}
}
\email{jiamang.wang@alibaba-inc.com}

\author{Han Zhu}
\affiliation{%
  \institution{Alibaba Group}
  \country{China}
}
\email{zhuhan.zh@alibaba-inc.com}

\author{Jian Xu}
\affiliation{%
  \institution{Alibaba Group}
  \country{China}
}
\email{xiyu.xj@taobao.com}

\author{Bo Zheng}
\affiliation{%
  \institution{Alibaba Group}
  \country{China}
}
\email{bozheng@alibaba-inc.com}

\author{Lin Qu}
\affiliation{%
  \institution{Alibaba Group}
  \country{China}
}
\email{xide.ql@taobao.com}

\renewcommand{\shortauthors}{Zong et al.}

\makeatletter

\renewcommand{\@titlenotes}{%
  \footnotetext[1]{%
    Corresponding author: Hua Zong
    (yuhuan.zh@alibaba-inc.com).
  }%
}

\let\RecISOriginalAuthorNotes\@authornotes

\renewcommand{\@authornotes}{%
  \setcounter{footnote}{1}%
  \RecISOriginalAuthorNotes
}

\let\RecISOriginalAuthorLayout\@mkauthors@iii

\newbox\RecISSavedTitle

\renewcommand{\@mkauthors@iii}{%

  \global\setbox\RecISSavedTitle=\copy\mktitle@bx

  \begingroup

    \def\@typeset@author@bx{%
      \gdef\@currentauthors{}%
      \gdef\@currentaffiliation{}%
    }%

    \RecISOriginalAuthorLayout

  \endgroup

  \global\let\and\@typeset@author@bx

  \global\setbox\mktitle@bx=\vbox{%

    \hsize=\textwidth

    \noindent
    \unvbox\RecISSavedTitle

    \par\medskip

    {%
      \centering
      \normalfont\large

      Hua Zong\textsuperscript{1,*\textdagger},
      Shuai Yang\textsuperscript{2,\textdagger},
      Zhengxiong Zhou\textsuperscript{1},
      Qingtao Zeng\textsuperscript{1},
      Jiawei Liu\textsuperscript{1},
      Hao Wan\textsuperscript{1},
      Hongbin Zhou\textsuperscript{1},
      Jiahao Wang\textsuperscript{1},
      Hao Zhang\textsuperscript{1},
      Bencheng Yan\textsuperscript{1},
      Yuheng Huang\textsuperscript{1},
      Huimin Yi\textsuperscript{1},
      Jiamang Wang\textsuperscript{1},
      Han Zhu\textsuperscript{1},
      Jian Xu\textsuperscript{1},
      Bo Zheng\textsuperscript{1},
      Lin Qu\textsuperscript{1}

      \par
    }

    \medskip

    {%
      \centering
      \normalfont\large

      \textsuperscript{1}Alibaba Group\par

      \textsuperscript{2}University of Chinese
      Academy of Sciences\par

      China\par
    }

    \medskip

    {%
      \centering
      \normalfont\normalsize

      \{yuhuan.zh,zhouzhengxiong.zzx,zengqingtao.zqt,
      lanling.ljw,wanhao.wh,hongbin.zhouhb,wjh177423,
      zh138764,huimin.yhm,jiamang.wang,zhuhan.zh,
      bozheng\}@alibaba-inc.com
      \par

      \{bencheng.ybc,hexing.hyh,xiyu.xj,xide.ql\}@taobao.com, yangshuai2022@iscas.ac.cn
      \par
    }

    \par\bigskip

  }%
}

\makeatother

\begin{abstract}

Modern recommender models scale sparse representations and dense computation together. They process thousands of heterogeneous feature fields, long user-behavior sequences, and trillion-scale Embedding parameters, while their dense networks increasingly use large Transformer architectures. The resulting execution mismatch creates fine-grained operations within stages, load imbalance and synchronization delays across devices, and GPU resource bubbles between stages. As models grow, these costs become a primary limit to scalable recommender training. Although existing systems improve Embedding caching and data prefetching, these inefficiencies still coexist in the GPU-centric workloads we study and jointly limit end-to-end throughput.

We present RecIS, a unified framework for GPU-centric recommender training. RecIS treats the complete path from input features to dense computation as one optimization target and coordinates three complementary mechanisms. First, adaptive fusion combines fragmented feature transformations with vectorized memory access. Second, a GPU Embedding engine uses hash-based sharding and hierarchical communication to balance request placement and eliminate redundant transfers; it also specializes aggregation for forward and backward index structures. Third, a multistage pipeline overlaps input preparation for future iterations with dense computation in the current iteration. Together, these mechanisms reorganize a fragmented sparse front end into a dataflow that continuously feeds dense computation.

We evaluate RecIS on representative industrial recommender models across distributed training platforms. Compared with PICASSO and TorchRec, RecIS reaches up to $2.62\times$ and $1.15\times$ their throughput, respectively, while reducing step latency by up to 61.5\% and 12.9\%. RecIS provides an efficient and scalable training system for large recommender models.

\end{abstract}


\maketitle
\hypersetup{pdfauthor={Hua Zong, Shuai Yang, Zhengxiong Zhou, Qingtao Zeng, Jiawei Liu, Hao Wan, Hongbin Zhou, Jiahao Wang, Hao Zhang, Bencheng Yan, Yuheng Huang, Huimin Yi, Jiamang Wang, Han Zhu, Jian Xu, Bo Zheng, Lin Qu}}

\section{Introduction}

Modern recommender training places two nearly opposite execution paradigms on the same GPU dataflow. The sparse front end processes thousands of heterogeneous feature fields, long behavior sequences, and trillion-scale embedding parameters through fragmented transformations, random lookups, and irregular communication~\cite{naumov2019deeplearningrecommendationmodel,zhang2021trainingefficiency,zhang2022picasso}. The dense back end is evolving toward billion-parameter Transformers and sparse mixture-of-experts models~\cite{vaswani2017attention,fedus2022switch,zhai2024hstu,yan2026lum} that require regular, continuous, and compute-intensive tensor pipelines. As both sides scale, the central systems challenge is no longer to accelerate sparse or dense components in isolation. It is to reorganize fragmented, skewed, and unevenly arriving sparse work into a stable stream that continuously feeds dense computation.

This mismatch creates a chain of bottlenecks whose effects compound along the dataflow. Thousands of variable-length transformations decompose input processing into small kernels and scattered memory accesses. Skewed feature accesses further create imbalanced embedding lookups and all-to-all communication, allowing the slowest device to determine the iteration time. Coarse iteration boundaries finally expose sparse-side tail latency to the dense network, leaving compute, memory, and communication resources idle in turn. In other words, task granularity determines sparse execution efficiency, device load determines its tail latency, and iteration scheduling determines whether that latency can be hidden by dense computation. Existing systems improve operator packing, embedding caching, or data prefetching~\cite{zhang2022picasso,ivchenko2022torchrec,zhang2024recflex,bagpipe2023}, but these mechanisms are generally designed around a particular stage or a fixed execution boundary. When the three problems occur together, accelerating one stage can simply move the bottleneck to the next point in the dataflow.

We present RecIS, a unified framework for GPU-centric recommender training. RecIS treats the path from raw features to dense computation as a schedulable dataflow and reorganizes sparse work at three connected levels:

\begin{itemize}
  \item \textbf{Across operators.} Adaptive fusion regularizes task granularity by grouping compatible transformations across fields into fewer CUDA kernels while preserving dependencies. Vectorized memory access reduces instruction overhead and improves bandwidth utilization.
  \item \textbf{Across devices.} Hash-based sharding distributes ownership of distinct IDs across workers in expectation. Hierarchical communication merges requests within each source node to reduce inter-node traffic, then deduplicates requests at each owner to reduce physical lookups. Index-aware aggregation reduces irregular memory access and atomic contention.
  \item \textbf{Across iterations.} The multistage pipeline overlaps future input preparation that is independent of parameter updates with current dense computation, using their complementary resource demands to hide sparse-side overhead.
\end{itemize}

Together, these mechanisms reduce sparse-stage overhead and overlap the remaining preparation work, allowing the sparse front end to feed dense computation more continuously. The source code of RecIS is publicly available at \url{https://github.com/alibaba/recis}.

\section{Background and Motivation}

\subsection{Feature-Preprocessing Bottlenecks}
\label{sec:preprocess}

GPU-centric training systems increasingly move feature preprocessing from CPUs to GPUs~\cite{zhang2022picasso,zhang2024recflex}. Offloading hashing, bucketization, and sequence truncation does not by itself ensure high throughput.

\begin{figure}[t]
  \centering
  \includegraphics[width=\linewidth]{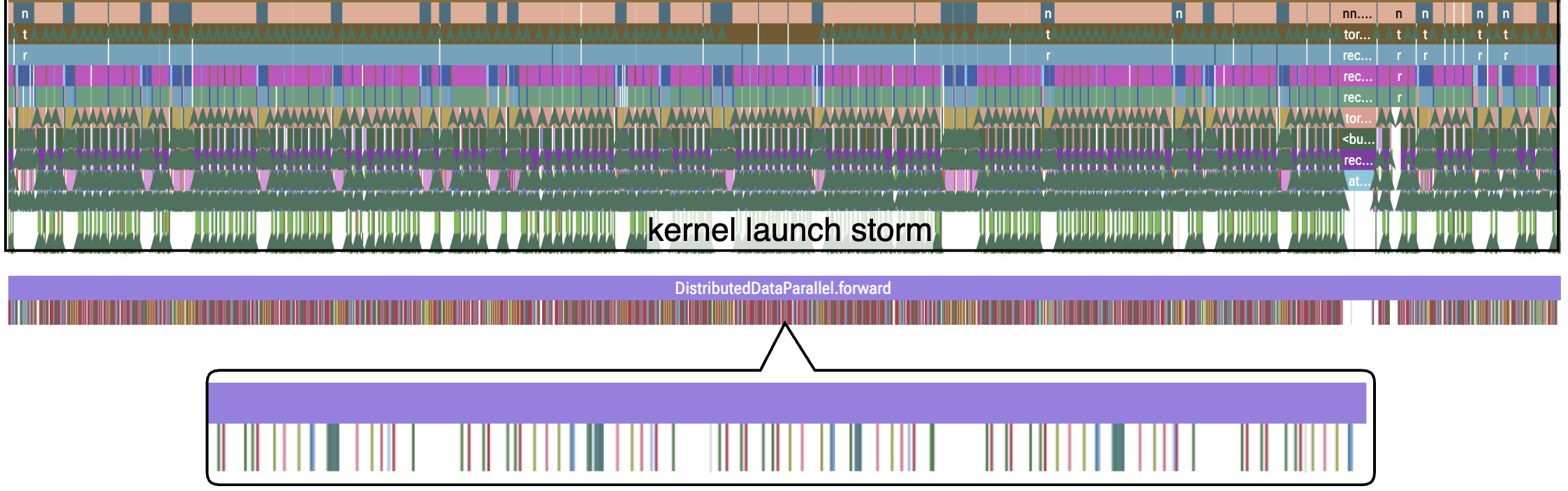}
  \caption{A profiling timeline exposes fragmented feature-preprocessing work and frequent kernel launches.}
  \Description{A GPU timeline contains many short preprocessing kernels separated by launch and scheduling gaps.}
  \label{fig:timeline}
\end{figure}

Production models contain thousands of heterogeneous feature fields, each with a chain of two to five preprocessing operations. Executing each field independently launches many short kernels, whose launch and scheduling overhead can exceed their useful computation (Figure~\ref{fig:timeline}). Fusion reduces these costs, but feature fields differ in both operator dependencies and sequence lengths. Small fusion groups leave substantial launch overhead, whereas overly coarse groups can introduce divergence and long-running tasks. Effective fusion must therefore account for both the operator structure and the distribution of input sizes.

Memory access presents a separate constraint. Hashing and modulo operations have low arithmetic intensity and are often limited by global-memory bandwidth~\cite{williams2009roofline}. Scalar accesses add memory instructions and can underutilize the bandwidth available to contiguous, aligned transfers.

\subsection{Repeated Requests and Embedding Imbalance}
\label{sec:embedding-motivation}

Embedding accesses are highly skewed. Popular IDs recur both within and across GPUs, amplifying lookup and communication work. Even a balanced distribution of stored parameters need not yield a balanced distribution of requests. Table-wise and contiguous row-wise sharding place parameters by table size or ID range, which need not match the access distribution of a particular batch~\cite{ivchenko2022torchrec,li2023fec,zhang2022picasso}. Unequal lookup and all-to-all traffic loads leave dense computation waiting for the slowest device.

Aggregation remains expensive after requests reach their owners. Forward outputs follow ordered compressed sparse row (CSR) segments defined by offsets, whereas backward contributions can target embedding rows in an arbitrary and highly concentrated order. Using the same element-wise atomic scatter for both paths introduces unnecessary forward atomics and contention at popular rows during backward execution. The embedding engine must therefore address request imbalance, duplicate accesses, and index-dependent aggregation costs together.

\subsection{Idle Time at Iteration Boundaries}
\label{sec:serial}

Training systems commonly organize execution around fixed iteration boundaries. Even with prefetching or local communication overlap, input I/O, feature processing, embedding operations, and dense computation often follow a largely sequential schedule. Our profiling shows that these stages stress different resources. Executing them in sequence leaves compute, memory, and I/O resources idle at different points in the iteration.

Some of this idle time comes from scheduling constraints rather than data dependencies. Work independent of parameter updates can begin before the current iteration finishes. However, prefetching a single batch is insufficient when preparation exceeds the available overlap window, and unrestricted background execution can interfere with dense kernels. Effective scheduling must expose more parallelism while preserving parameter consistency and controlling contention.

\section{RecIS System Design}

RecIS jointly optimizes feature preprocessing, distributed embedding operations, and pipeline scheduling. Each component reduces overhead that would otherwise delay the next stage of the sparse-to-dense dataflow.

\subsection{Fused and Vectorized Feature Processing}
\label{sec:feature-engine}

FeatureEngine addresses kernel-launch overhead, work imbalance across feature fields, and inefficient scalar memory access.

\paragraph{Dependency-aware fusion.} Feature fields differ in both their transformation chains and input lengths. Small fusion groups retain substantial launch overhead, whereas overly coarse groups can introduce divergence and long-running tasks. FeatureEngine therefore fuses independent operations of the same category while preserving the dependencies of each feature.

\begin{figure}[t]
  \centering
  \includegraphics[width=\linewidth]{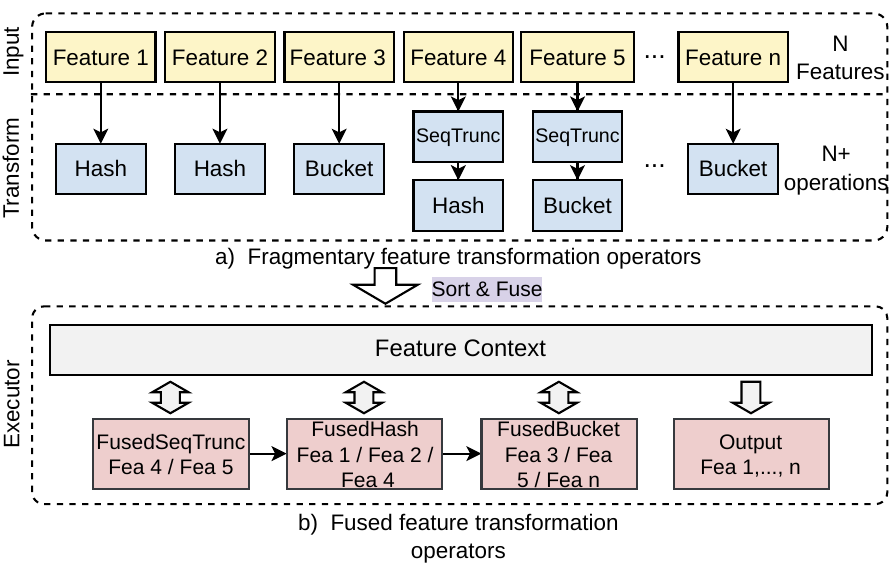}
  \caption{FeatureEngine preserves operator dependencies while fusing compatible transformations across feature fields.}
  \Description{Per-feature transformation chains are reordered by dependency and combined into shared kernels for compatible operator categories.}
  \label{fig:feature-fusion}
\end{figure}

The transformations in Figure~\ref{fig:feature-fusion} include bucketization, modulo, hashing, sequence truncation, and masking. The scheduler first topologically orders these operations. For example, \texttt{SeqTrunc} must finish before \texttt{Hash} for Feature 4. Independent operations from other fields can join the same fusion group when they are of a compatible category. Fusion thus shares execution across fields without reordering a dependent transformation ahead of its input.

For $F$ feature fields, let $d_i$ be the number of operations on field $i$ and $\mathcal{G}$ the resulting fusion groups. The kernel-launch counts satisfy
\begin{equation}
  K_{\mathrm{RecIS}}=|\mathcal{G}|\le K_{\mathrm{base}}
  =\sum_{i=1}^{F}d_i.
  \label{eq:fusion-launches}
\end{equation}
Each group replaces multiple launches when fields share compatible transformations; otherwise, the inequality allows equality.

\paragraph{Thread mapping for skewed inputs.} Fusion exposes more work to each kernel, but equal treatment of feature fields would not imply equal work for threads. Long behavior sequences can contain many more elements than short categorical fields. Static assignment by field would carry this skew into the fused kernel, leaving some threads idle while others process long sequences.

FeatureEngine flattens elements across fields into one logical index space and partitions it into equally sized tiles. A grid-stride mapping distributes these tiles to threads, which use feature-boundary indices to locate the corresponding input intervals. Work is therefore divided by element count rather than by the number of fields. This mapping reduces idle threads and limits the execution tail of long inputs without changing the transformation assigned to each element.

\paragraph{Vectorized memory access.} Fusion does not by itself eliminate memory-instruction overhead. Hashing and modulo have low arithmetic intensity and are often limited by memory bandwidth~\cite{williams2009roofline}. Element-wise operators use fixed-width vector loads and stores where dependencies, alignment, and boundaries permit. Packed kernels process groups of elements; \texttt{select\_pack\_size} selects the largest valid pack at compile time, with up to 16 bytes accessed per thread per memory instruction. Smaller valid packs accommodate stricter alignment or boundary constraints. Contiguous, aligned transfers reduce instruction count and improve effective bandwidth.

\textbf{In summary,} FeatureEngine regularizes feature preprocessing through these three complementary mechanisms, consolidating thousands of heterogeneous transformation chains into a small set of regular, high-throughput GPU kernels. This design improves data preparation for subsequent embedding lookups and pipelined execution across stages.

\subsection{Embedding Engine with Hierarchical Communication}
\label{sec:embedding-engine}

\paragraph{Hash-based sharding.} Balanced storage need not imply balanced requests: table-wise and contiguous row-wise placement can concentrate popular tables or ID ranges. RecIS hashes IDs to spread ownership across workers in expectation, reducing this dependence on table size or ID range. Sharding distributes distinct IDs; repeated requests for the same ID still require deduplication.

\begin{figure}[t]
  \centering
  \includegraphics[width=\linewidth]{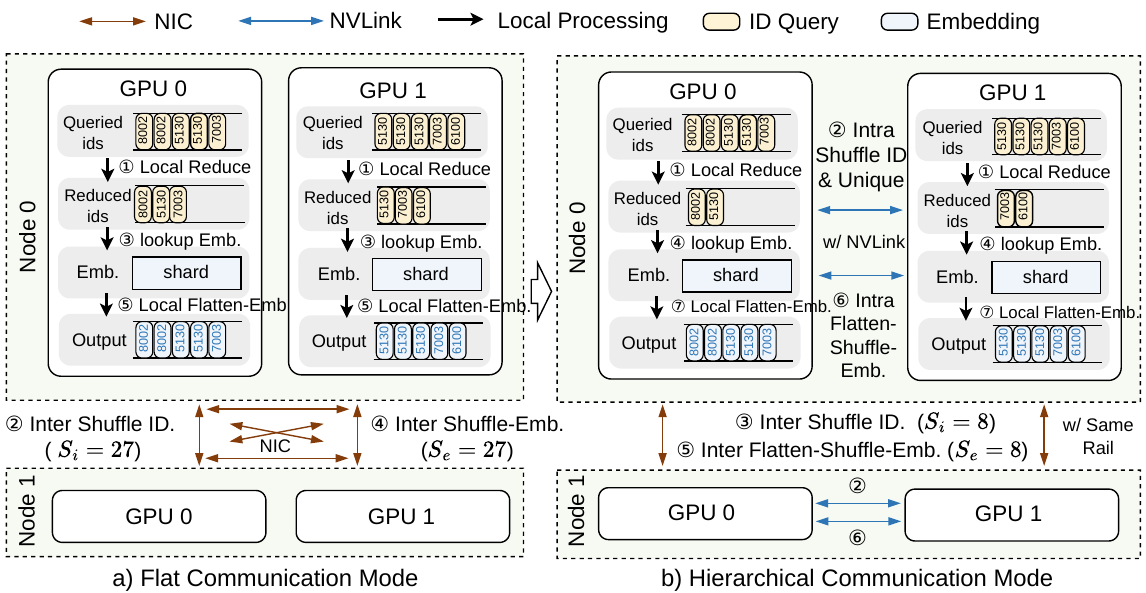}
  \caption{Flat and hierarchical embedding communication. $S_i$ counts IDs sent across nodes; $S_e$ counts embedding vectors returned. The reduced IDs in (b) are the result of steps~\step{1} and~\step{2}; step~\step{3} also deduplicates requests at the owner.}
  \Description{The Flat path exchanges the unique requests from each GPU directly with owners. The hierarchical path merges requests over NVLink, exchanges them along corresponding inter-node rails, and expands returned vectors within each source node. Both paths restore the same per-GPU outputs.}
  \label{fig:embedding-communication}
\end{figure}

\paragraph{Hierarchical communication.} Flat communication (Figure~\ref{fig:embedding-communication}(a)) deduplicates IDs per GPU but still sends separate requests for IDs shared across GPUs. RecIS merges these requests within each source node over NVLink, routes them to their owners over same-rail links, and distributes each returned vector locally to the requesting GPUs.

Figure~\ref{fig:embedding-communication}(b) shows the forward path. Each deduplication stage retains an inverse map for restoring the original requests.
\step{1} \emph{Local Reduce-ID:} each GPU removes duplicate IDs from its local requests.
\step{2} \emph{Intra Shuffle-Reduce-ID:} NVLink routes requests to the source-node GPU with the same local GPU index as the owner, where duplicates from different GPUs are merged.
\step{3} \emph{Inter Shuffle-Reduce-ID:} same-rail groups connect corresponding GPUs across nodes and route remote requests to their owners, which merge duplicates from all source nodes.
\step{4} \emph{Lookup Embed:} each owner performs one lookup per unique ID in its local shard.
\step{5} \emph{Inter Flatten-Shuffle-Embed:} the inverse map from step~\step{3} expands responses for return over the same rails, with one vector per ID per remote source node.
\step{6} \emph{Intra Flatten-Shuffle-Embed:} the inverse map from step~\step{2} expands vectors for NVLink delivery to the requesting GPUs.
\step{7} \emph{Local Flatten-Embed:} the inverse map from step~\step{1} restores the original request order and multiplicity.

To compare traffic, consider one embedding table and the same batch, sharding, and GPU-local deduplication in both modes. There are $N$ nodes with $M$ GPUs each. Let $I_{n,m}$ be the set of unique IDs requested by GPU $m$ on node $n$ whose owners reside on other nodes, and let $I_n=\bigcup_{m=0}^{M-1}I_{n,m}$. Counting each logical inter-node transfer once, with one vector returned per request, gives
\begin{equation}
\begin{aligned}
S_i^{\mathrm{flat}}=S_e^{\mathrm{flat}}
&=\sum_{n=0}^{N-1}\sum_{m=0}^{M-1}|I_{n,m}|,\\
S_i^{\mathrm{hier}}=S_e^{\mathrm{hier}}
&=\sum_{n=0}^{N-1}|I_n|.
\end{aligned}
\label{eq:communication-counts}
\end{equation}
Let $G_n(x)=|\{m\mid x\in I_{n,m}\}|$ be the number of GPUs on node $n$ requesting ID $x$. The reduction in each type of transfer is
\begin{equation}
\begin{aligned}
\Delta S
&=S_i^{\mathrm{flat}}-S_i^{\mathrm{hier}}
 =S_e^{\mathrm{flat}}-S_e^{\mathrm{hier}}\\
&=\sum_{n=0}^{N-1}\sum_{x\in I_n}\bigl(G_n(x)-1\bigr)\ge 0.
\end{aligned}
\label{eq:communication-savings}
\end{equation}
Thus, step~\step{2} replaces $G_n(x)$ inter-node requests and returned vectors with one of each. Owner-side deduplication in step~\step{3} reduces physical lookups, but cannot remove transfers that have already occurred.

\paragraph{Index-aware segment reduction.}
Restoring gathered embeddings to the original ragged layout requires weighted segment aggregation. Let $\mathbf{E}\in\mathbb{R}^{U\times D}$ denote the unique embedding matrix, $\mathbf{o}$ the CSR offsets defining $S$ segments, $\mathbf{i}$ the reverse indices mapping $N$ contributions to embedding rows, and $\mathbf{w}$ their weights. Although forward and backward share the same reduction semantics, their destination indices exhibit different access patterns, motivating specialized GPU kernels. Algorithm~\ref{alg:index_segment} outlines their execution workflows.

\textit{\textbf{Forward.}}
The CSR offsets define a contiguous contributor range $[o_s,o_{s+1})$ for each output segment $s$. RecIS directly traverses these ranges and computes $\mathbf{Y}_{s,:}=\sum_{r=o_s}^{o_{s+1}-1} w_r\mathbf{E}_{i_r,:}$. Each thread owns one or more non-overlapping feature packs of an output segment, gathers the corresponding embedding values, and accumulates the weighted sum in registers. Once the segment is processed, the thread commits its result through a vectorized store. Even when embedding IDs repeat within a segment, all contributions to each output element are handled by its unique writer. Empty segments produce zero. This ownership-based mapping eliminates global atomic operations and their associated read-modify-write traffic.

\textit{\textbf{Backward.}}
In contrast, backward scatters gradients to embedding rows indexed by $\mathbf{i}$, whose entries may appear in arbitrary order and contain highly frequent duplicates. Each contributor $r$ generates $w_r\mathbf{G}_{s(r),:}$, where $s(r)$ is its forward segment ID and $\mathbf{G}$ is the output gradient. RecIS assigns a group of GPU lanes to each contributor row. Lanes cooperatively load contiguous feature packs, then redistribute values through shuffle operations to enable coalesced feature-wise updates. Within each wave, row groups with identical destination IDs are identified and their contributions are reduced using lane shuffles. Only the leader row group performs atomic additions for the merged feature values. For $k$ contributions targeting the same row within a wave, this reduces the number of scalar atomic updates from $kD$ to $D$. Contributions targeting the same row across different waves or thread blocks are still combined through global atomics, preserving correctness without sorting destination indices.

\begin{algorithm}[t]
\caption{Index-aware Segment Reduction}
\label{alg:index_segment}
\footnotesize
\begin{algorithmic}[1]

\State \textbf{Forward: CSR-guided reduction}

\ForAll{output feature packs $(s,p)$ \textbf{in parallel}}
    \State $\mathbf{a} \gets \mathbf{0}$
    \For{$r=o_s$ \textbf{to} $o_{s+1}-1$}
        \State $\mathbf{a} \gets \mathbf{a}
               + w_r \mathbf{E}[i_r,p]$
    \EndFor
    \State $\mathbf{Y}[s,p] \gets \mathbf{a}$
    \Comment{Unique writer}
\EndFor

\Statex
\State \textbf{Backward: Wave-level duplicate merging}

\State Initialize $\nabla\mathbf{E} \gets 0$

\ForAll{GPU waves $\omega$ \textbf{in parallel}}
    \State Assign one lane group to each contributor $r$
    \State Load $w_r\mathbf{G}[s(r),:]$ into lanes
    \State Match identical destination IDs $i_r$

    \ForAll{distinct destination IDs $u$ in $\omega$}
        \State $\mathbf{a}_u \gets
            \sum_{r\in\omega:\,i_r=u}
            w_r\mathbf{G}[s(r),:]$
        \State Elect one leader row group
        \ForAll{features $d$ assigned to leader lanes}
            \State $\mathrm{atomicAdd}(
                \nabla\mathbf{E}[u,d],a_{u,d})$
        \EndFor
    \EndFor
\EndFor

\end{algorithmic}
\end{algorithm}

\subsection{Multistage Pipeline Execution}
\label{sec:pipeline-engine}

Fixed iteration boundaries can serialize input preparation with dense computation even after the feature and embedding engines reduce sparse-stage costs. A largely sequential schedule leaves different resources idle at different times. The multistage pipeline in RecIS begins independent work earlier while preserving the dependencies of parameter access and updates.

\paragraph{Resource demands and dependencies.} Two observations guide scheduling. First, stages stress different resources. Input I/O uses host processing and interconnect bandwidth; feature transformations and request formatting mainly consume memory bandwidth; physical embedding lookup accesses device memory; and dense forward and backward passes mainly use Tensor Core computation. Overlapping these partly complementary demands can use otherwise idle capacity.

Second, iteration boundaries can impose waits beyond those required by training semantics. RecIS restricts asynchronous cross-iteration preparation to modules that do not participate in backpropagation and do not depend on parameter values updated by the current iteration. Input loading, feature transformation, and ID request formatting satisfy these conditions and can proceed before the current update finishes. Physical embedding lookup, dense forward and backward computation, and all parameter updates remain in the main training workflow with their original ordering dependencies. In particular, preparation does not prefetch embedding values. The next iteration reads them only after the preceding updates on which it depends have completed. This separation preserves parameter-access and update semantics while allowing independent preparation to overlap training.

\setlength{\emergencystretch}{1.5em}
\paragraph{Stage decomposition and scheduling.} We divide each iteration into six stages: \texttt{SPARSE\_FORWARD}, \texttt{DENSE\_FORWARD}, \texttt{METRICS\_OP}, \texttt{BACKWARD}, \texttt{OPTIM\_STEP}, and \texttt{PRE\_FORWARD}. \texttt{SPARSE\_FORWARD} performs physical embedding lookups. \texttt{PRE\_FORWARD} contains input I/O, feature transformations, and embedding request formatting and partitioning. Stage descriptors distinguish this preparation from the parameter-dependent training path.
\par
\setlength{\emergencystretch}{0pt}

\begin{figure}[t]
  \centering
  \includegraphics[width=\linewidth]{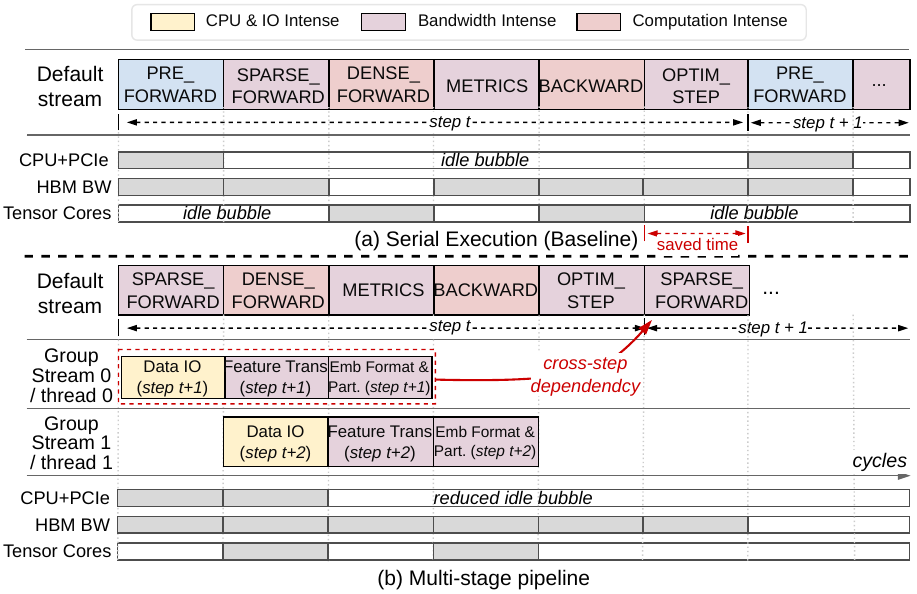}
  \caption{Multistage scheduling overlaps future input preparation with current dense computation while preserving dependencies.}
  \Description{A comparison of serial and overlapping schedules shows future-iteration preparation on a Group Stream alongside current computation on the default stream.}
  \label{fig:pipeline}
\end{figure}

Figure~\ref{fig:pipeline} shows the resulting schedule. Iteration $t$ follows its dependency order on the default stream. Independent preparation for subsequent iterations runs in a separate asynchronous workflow, the \emph{Group Stream}, alongside compatible current-iteration stages. For example, input loading, feature transformations, and request formatting for iteration $t+1$ can overlap \texttt{DENSE\_FORWARD} of iteration $t$. The prepared requests then feed \texttt{SPARSE\_FORWARD} when its parameter dependencies are satisfied. The overlap pattern is configurable to suit the dependencies and resource demands of each model.

\paragraph{Preparation depth and contention.} Preparation time varies with the model and input. Prefetching a single batch is insufficient when this work exceeds the available overlap window: the unhidden remainder still delays the next iteration. RecIS supports multiple preparation stages that distribute work across threads and asynchronous streams. Figure~\ref{fig:pipeline} illustrates preparation for successive iterations in separate Group Stream workflows, allowing preparation to extend beyond the overlap window of a single iteration.

More concurrency is useful only when contention remains manageable. Background kernels can compete with dense kernels for execution resources and memory bandwidth even when the dense computation is dominated by Tensor Core GEMMs. Asynchronous streams therefore use the conservative \texttt{StreamPriority.DEFAULT} policy to limit interference while exploiting otherwise idle capacity. Stage count and overlap topology provide controls for different workloads; the appropriate schedule depends on preparation time relative to the available overlap window.

The pipeline hides preparation work that remains after feature and embedding optimization, while retaining the order required for parameter-dependent computation.

\section{Evaluation}
\label{sec:evaluation}

We evaluate end-to-end throughput and step latency, then examine feature fusion, hierarchical request merging, index-aware segment reduction, and multistage scheduling.

\subsection{Experimental Setup}
\label{sec:eval-setup}

We evaluate three industrial recommender models on distributed training infrastructure: \textit{Ad} and \textit{Lazada Ad} for advertising ranking, and \textit{Search} for search ranking. Table~\ref{tab:model-specifications} summarizes their input features, logical embedding tables, embedding dimensions, and per-device batch sizes. \textit{Ad} runs on 96 PPUs, \textit{Search} on 64 NVIDIA H20 GPUs, and \textit{Lazada Ad} on 8 NVIDIA L20 GPUs. For each model, all compared systems use the same hardware and training configuration.

We compare RecIS with PICASSO~\cite{zhang2022picasso} and TorchRec~\cite{ivchenko2022torchrec}. For each model, all three frameworks use the same hardware, model, batch size, numerical precision, and optimizer configuration, including learning rates. End-to-end performance values are averaged across repeated runs to reduce the influence of run-to-run variability

\begin{table}[t]
  \caption{Statistics of the evaluated recommender models.}
  \label{tab:model-specifications}
  \centering
  \small
  \setlength{\tabcolsep}{4pt}
  \resizebox{\columnwidth}{!}{%
    \begin{tabular}{lrrrlcc}
      \toprule
      Model &
      \shortstack{Input\\features} &
      \shortstack{Logical emb.\\tables} &
      \shortstack{Batch /\\device} &
      \shortstack{Embedding\\dimensions} &
      \shortstack{Device\\Type} &
      \shortstack{Nodes $\times$\\devices/node} \\
      \midrule
      \textit{Ad} &
      879 & 519 & 1,200 & 8--128 & PPU 810e & $6\times16$ \\

      \textit{Search} &
      589 & 398 & 256 & 4--128 & H20 & $8\times8$\\

      \textit{Lazada Ad} &
      367 & 73 & 1,024 & 4--416 & L20 & $1\times8$\\
      \bottomrule
    \end{tabular}%
  }
\end{table}

\subsection{End-to-End Performance}
\label{sec:e2e-performance}

\begin{figure}[t]
  \centering
  \includegraphics[width=\linewidth]{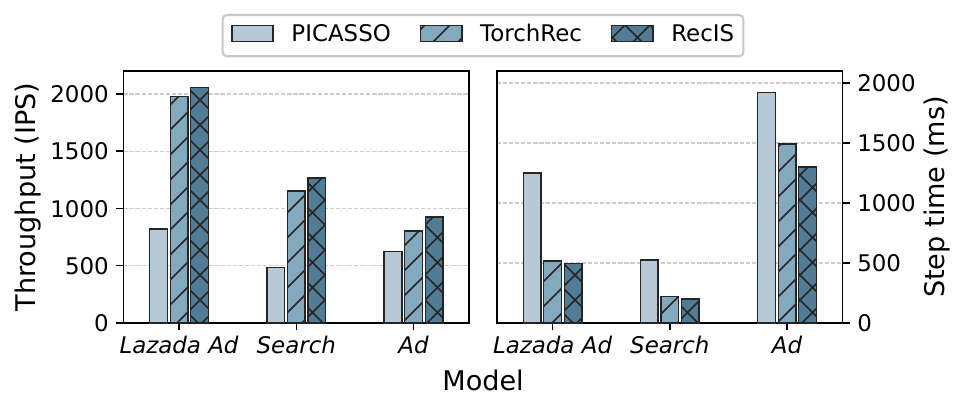}
  \caption{End-to-end throughput per GPU and step latency for PICASSO, TorchRec, and RecIS on three industrial workloads.}
  \Description{Two bar-chart panels compare throughput in instances per second and step latency in milliseconds on Lazada Ad, Search, and Ad.}
  \label{fig:e2e-performance}
\end{figure}

Figure~\ref{fig:e2e-performance} reports throughput per GPU and step latency. RecIS achieves the highest throughput and lowest latency on all three workloads. It delivers $1.04$--$1.15\times$ the throughput of TorchRec with 3.9\%--12.9\% lower step latency, and $1.48$--$2.62\times$ the throughput of PICASSO with 32.3\%--61.5\% lower latency. These results establish the end-to-end benefit of the combined system; the component experiments examine the underlying reductions in sparse work and exposed latency.

RecIS reaches 924 IPS at 1,299\,ms per step on \textit{Ad}, 1,268 IPS at 202\,ms on \textit{Search}, and 2,058 IPS at 498\,ms on \textit{Lazada Ad}. These workloads use different deployment scales, so comparisons use a fixed training configuration for each model. The maximum gain over PICASSO occurs on \textit{Search}, whereas the largest relative gain over TorchRec occurs on \textit{Ad}.

\subsection{Training Loss and AUC}
\label{sec:training-behavior}

To evaluate whether RecIS's optimizations
affect model accuracy, we compare its training behavior
against TorchRec using the \textit{Search} model.
Both systems are trained from scratch for 5,000 steps
under identical hyperparameters, including a per-device
batch size of 256, a sparse learning rate of 0.01,
and a dense learning rate of $10^{-5}$.
TorchRec runs on 32 NVIDIA H20 GPUs, while RecIS
runs on 32 PPU 810e accelerators.

\begin{figure}[t]
  \centering
  \includegraphics[width=\linewidth]{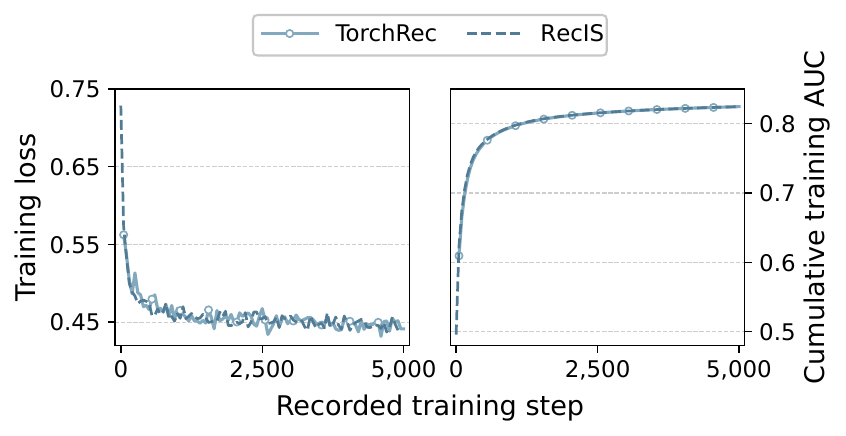}
  \caption{Training loss and cumulative AUC of the \textit{Search} model under TorchRec and RecIS.}
  \Description{Two line charts compare TorchRec on 32 H20 GPUs with RecIS on 32 PPU 810e accelerators. Training loss decreases to approximately 0.441 in both runs, while cumulative training AUC increases to approximately 0.824 at their respective last recorded points. TorchRec has no recorded step-zero point, and RecIS has no recorded step-5000 point.}
  \label{fig:training-convergence}
\end{figure}

Figure~\ref{fig:training-convergence} compares the
training loss and cumulative AUC of both systems.
The loss is averaged across all 32 ranks at each
sampled step, while the AUC is computed from globally
aggregated statistics accumulated throughout training.
Both systems exhibit similar training trends, with
decreasing loss and increasing AUC.
At their respective final recorded points, TorchRec
and RecIS achieve training losses of 0.441442 and
0.441274, and cumulative AUCs of 0.824599 and
0.824424, respectively.
These closely matched results indicate that RecIS's
optimizations preserve training accuracy without
observable degradation compared to TorchRec.

\subsection{Component Analysis}
\label{sec:design-evaluation}

\begin{figure}[t]
  \centering
  \includegraphics[width=\linewidth]{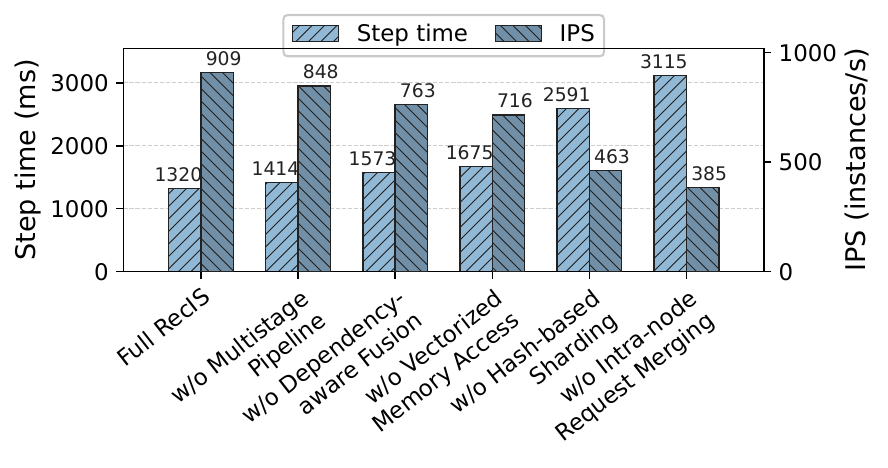}
  \caption{Single-node cumulative ablation on \textit{Ad} with eight H20 GPUs. From left to right, each configuration additionally disables the named optimization. Request merging denotes intra-node ID deduplication.}
  \Description{Bars show step latency and throughput for full RecIS and cumulative removal of multistage scheduling, dependency-aware fusion, vectorized access, hash-based sharding, and intra-node request merging.}
  \label{fig:ablation}
\end{figure}

\textbf{Cumulative ablation.} We evaluate successive configurations of \textit{Ad} on a single server with eight H20 GPUs (Figure~\ref{fig:ablation}). Starting from the configuration with all studied optimizations disabled, intra-node request merging and hash-based sharding raise throughput from 385.19 to 716.26 IPS and reduce step latency from 3,115.36 to 1,675.38\,ms, a 46.2\% reduction. Vectorized access and dependency-aware fusion further increase throughput to 848.37 IPS and reduce latency to 1,414.48\,ms. With these optimizations enabled, the multistage pipeline removes another 94\,ms and raises throughput by 7.1\%, reaching 908.76 IPS at 1,320.48\,ms per step. Overall, RecIS achieves $2.36\times$ the baseline throughput and reduces latency by 57.6\%.

\textbf{Dependency-aware fusion.} Table~\ref{tab:operator-count} reports preprocessing operator invocations before and after cross-field fusion. Counts fall from 1,351 to 79 on \textit{Ad}, from 1,071 to 158 on \textit{Search}, and from 688 to 136 on \textit{Lazada Ad}: reductions of 94.15\%, 85.25\%, and 80.23\%, respectively. Across the three models, fusion reduces 3,110 invocations to 373, an 88.01\% reduction. Fewer invocations reduce the launch and scheduling overhead addressed by FeatureEngine.

\begin{table}[t]
  \caption{Preprocessing operator invocations before and after cross-field fusion.}
  \label{tab:operator-count}
  \centering
  \small
  \begin{tabular}{lrrr}
    \toprule
    Model & Baseline & RecIS & Reduction \\
    \midrule
    \textit{Ad} & 1,351 & 79 & 94.15\% \\
    \textit{Search} & 1,071 & 158 & 85.25\% \\
    \textit{Lazada Ad} & 688 & 136 & 80.23\% \\
    \bottomrule
  \end{tabular}
\end{table}

\textbf{Hierarchical communication.}
To evaluate the effectiveness of hierarchical communication,
we compare its communication traffic and overhead against
Flat communication using an advertising model deployed on
16 nodes with 128 accelerators. Both approaches use identical
configurations except for the communication strategy.
Table~\ref{tab:hier-communication-cost} reports the average
per-step communication costs across all ranks for three
representative embedding tables.
Compared with Flat, Hierarchical reduces inter-node embedding
traffic by 59.34\%, 72.37\%, and 31.71\%, respectively,
with corresponding retained ID fractions of 40.71\%,
27.65\%, and 68.36\%.

These reductions are achieved by merging redundant embedding
requests within each source node before inter-node communication
(Eq.~\ref{eq:communication-savings}).
However, Hierarchical introduces additional overhead from
intra-node embedding redistribution and ID partitioning.
Consequently, its performance benefit depends on whether
the inter-node communication savings outweigh these
additional costs.

The timing breakdown further illustrates this trade-off.
For the FP32 embedding table with dimension 32,
Hierarchical reduces inter-node communication time from
76.695\,ms to 30.487\,ms, while introducing 5.405\,ms
of intra-node embedding communication and 5.262\,ms
of additional ID partitioning.
In contrast, for the smaller FP32 table with dimension 16,
the combined embedding communication time is 1.291\,ms,
only 0.740\,ms lower than Flat, while the additional ID
partitioning takes 0.851\,ms.
These results demonstrate that reducing inter-node traffic
alone does not guarantee a performance improvement,
as the additional intra-node processing overhead may
offset the communication savings.
The reported component times do not directly represent
critical-path latency or end-to-end training time.

\begin{figure}[t]
  \centering
  \includegraphics[width=\linewidth]{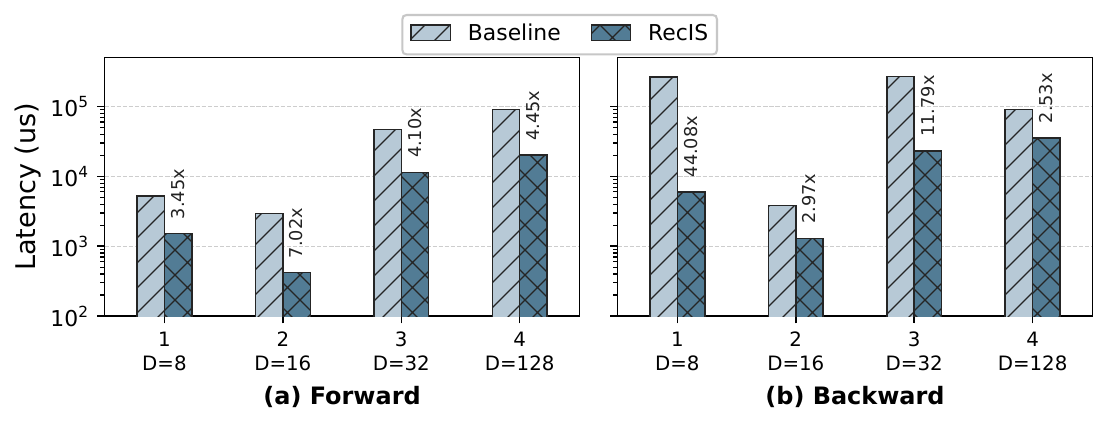}
  \caption{Index-aware segment reduction on AMD MI308X. $D$ is the embedding dimension; annotations show speedup over the original kernels.}
  \Description{Forward and backward latency is compared for the original and specialized kernels on workloads with embedding dimensions of 8, 16, 32, and 128.}
  \label{fig:segment-reduction}
\end{figure}

\textbf{Index-aware segment reduction.} We compare the original and specialized aggregation kernels on AMD MI308X across four workloads (Figure~\ref{fig:segment-reduction}). The forward kernel uses CSR segments to assign unique writers; the backward kernel locally merges contributions to the same row. Forward latency decreases by 71.0\%--85.8\% and backward latency by 60.5\%--97.7\%, with speedups of up to $7.02\times$ and $44.08\times$, respectively. Exploiting the index structure substantially reduces aggregation overhead in both directions.

\begin{table}[t]
  \caption{Comparison of communication traffic and overhead
between Flat and Hierarchical embedding communication.
All measurements are per-step averages across ranks.}
  \label{tab:hier-communication-cost}
  \centering
  \small
  \resizebox{\columnwidth}{!}{%
    \begin{tabular}{lrrr}
      \toprule
      Table & \texttt{1} & \texttt{2} & \texttt{3} \\
      Type / dimension & FP32 / 16 & FP32 / 32 & int8 / 128 \\
      \midrule
      Flat emb traffic (GB) & 2.976 & 215.590 & 193.013 \\
      Hier inter emb traffic (GB) & 1.210 & 59.568 & 131.806 \\
      Retained IDs (\%) & 40.71 & 27.65 & 68.36 \\
      \midrule
      Flat emb time (ms) & 2.031 & 76.695 & 69.227 \\
      Hier intra emb time (ms) & 0.088 & 5.405 & 5.003 \\
      Hier inter emb time (ms) & 1.203 & 30.487 & 53.622 \\
      Hier extra processing (ms) & 0.851 & 5.262 & 5.487 \\
      \bottomrule
    \end{tabular}%
  }
\end{table}

\section{Related Work}

{Wide \& Deep}, {DeepFM}, {xDeepFM}, and {DCN V2} model feature interactions~\cite{cheng2016widedeep,guo2017deepfm,lian2018xdeepfm,wang2021dcnv2}, while {DIN} and {DIEN} model user interests from behavior sequences~\cite{zhou2018deepinterestnetwork,zhou2019dien}. To support these workloads, TensorFlow and PyTorch provide general execution primitives~\cite{abadi2016tensorflow,paszke2019pytorch}, while DeepRec, HugeCTR, TorchRec, and PERSIA specialize in large-scale recommendation training~\cite{alibaba_deeprec,wang2022hugectr,ivchenko2022torchrec,lian2022persia}.

Feature heterogeneity motivates tailored GPU execution. RecFlex adapts schedules to heterogeneous features~\cite{zhang2024recflex}. RecIS groups compatible transformations across fields while preserving dependencies, using fusion and vectorized access to reduce preprocessing overhead.

Embedding optimizations target execution, storage, and communication. Kwon and Rhu optimize GPU embedding execution~\cite{kwon2022gpuscratch}, while EL-Rec reduces storage via tensor-train representations~\cite{wang2022elrec}. ScaleFreeCTR and HET cache embeddings to reduce remote access~\cite{guo2021scalefreectr,miao2021het}; FEC combines embedding tiering with prefetching to reduce communication overhead~\cite{li2023fec}. RecIS merges repeated IDs across GPUs in each source node before inter-node exchange, reducing both transmitted requests and returned vectors.

Scheduling complements these optimizations by overlapping preparation and training. RAP co-schedules GPU preprocessing and training~\cite{wang2024rap}, while Bagpipe uses future minibatches to prefetch embeddings~\cite{bagpipe2023}. PICASSO coordinates embedding-operation packing, interleaved execution, and hierarchical caching~\cite{zhang2022picasso}. GPipe and PipeDream pipeline model stages across devices~\cite{huang2019gpipe,narayanan2019pipedream}. RecIS pipelines preparation across iterations: parameter-independent feature processing and request preparation overlap current dense computation, while parameter accesses and updates retain their required ordering.

\section{Conclusion}

RecIS coordinates three aspects of GPU-centric recommender training: operator granularity, embedding request distribution, and execution order. Adaptive fusion and vectorization reduce preprocessing overhead; sharding, hierarchical communication, and index-aware aggregation reduce embedding costs; and multistage scheduling overlaps future preparation with current dense computation. Across the evaluated workloads, RecIS achieves up to $1.15\times$ the throughput of TorchRec and $2.62\times$ that of PICASSO, with step-latency reductions of up to 12.9\% and 61.5\%, respectively.

\section{Acknowledgments}

This work is supported by Alibaba Group. We thank Chuan Wang, Han Wu, Kui Ma, Yuejie Lei, Feifan Yang, Bin Liu, Zhangming Chan, Ziru Xu, and Yong Bai for useful pointers regarding the writing of this paper. We appreciate Qi Wang, Yiwen Hu, Hechen Sun, Mingjie Liu, Yihan Xian, Zhensong Yan, Houyuan Xiang, Wenjie Guo, Shihang Yan, Ning Zhang, and Yingyu Ji for implementing the key components of the training infrastructure. We also thank Lei Wu, Tianhao Shen, Fuwei Yang, and ChunHung Wang from AMD for their technical support and assistance in optimizing core components of the framework.

\bibliographystyle{ACM-Reference-Format}
\bibliography{recis}

\end{document}